# Integrated Fast Ignition Simulation of Cone-guided Target with Three Codes

H. Sakagami[1)], T. Johzaki[2)], H. Nagatomo[2)], and K. Mima[2)]

[1)]*Computer Engineering, University of Hyogo, 2167 Shosha, Himeji, Hyogo 671-2201 Japan*

[2)]*Institute of Laser Engineering, Osaka University, 2-6 Yamadaoka, Suita, Osaka 565-0871 Japan*

*email: sakagami@eng.u-hyogo.ac.jp*

**Abstract.** It was reported that the fuel core was heated up to ~ 0.8 [keV] in the fast ignition experiments with cone-guided targets, but they could not theoretically explain heating mechanisms and achievement of such high temperature. Thus simulations should play an important role in estimating the scheme performance, and we must simulate each phenomenon with individual codes and integrate them under the Fast Ignition Integrated Interconnecting code project. In the previous integrated simulations, fast electrons generated by the laser-plasma interaction were too hot to efficiently heat the core and we got only 0.096 [keV] temperature rise. Including the density gap at the contact surface between the cone tip and the imploded plasma, the period of core heating became longer and the core was heated by 0.162 [keV], ~69% higher increment compared with ignoring the density gap effect.

## 1. Introduction

The fast ignition [1] approach to inertial fusion energy has been investigated by experiments with cone-guided targets [2,3], because this scheme is expected to reduce criteria for fuel burning. Simulations should play an important role in estimating the scheme performance, and the scientifically exciting new physics of the fast ignition should be self-consistently described in computations. Thus we must consider 1) overall fluid dynamics of the implosion, 2) laser-plasma interaction and super fast electron generation, and 3) super fast electron energy deposition within the core. It is, however, impossible to simulate all phenomena with a single simulation code, and we must simulate each phenomenon with individual codes and integrate them. Recently, we have started the Fast Ignition Integrated Interconnecting code ($FI^3$) project [4]. Under this project, the ALE (Arbitrary Lagrangian Eulerian) hydro code (PINOCO: Precision Integrated implosion Numerical Observation COde) [5], the collective PIC code (FISCOF1: Fast Ignition Simulation code with COllective and Flexible particles) [4], and the relativistic Fokker-Planck code (RFP: Relativistic Fokker-Planck code) [6] collaborate each other with data transfer via the computer networks. Since communication among these codes is very straightforward, we have designed a lightweight TCP/IP based protocol, Distributed Computing Collaboration Protocol (DCCP), to transfer data between codes [7].

Typical scenario in $FI^3$ project is summarized as follows: PINOCO computes implosion dynamics, including the absorption of a main implosion laser. FISCOF1 obtains density profile at the maximum compression from PINOCO and introduces plasma corresponding to that profile into the PIC system.

Then an ultrahigh intense ignition laser is launched into the plasma, and we can simulate laser-plasma interactions under realistic conditions in PIC simulations. FISCOF1 computes generation of fast electrons by the laser-plasma interaction, and regularly sends distribution functions of fast electrons to RFP. RFP receives the distribution functions, and treats them as a time dependent source term of fast electrons. RFP also gets the bulk plasma profile from PINOCO, and simulates core heating and fusion burning processes under obtained conditions [8]. Code integration diagram and physical quantities that should be communicated between codes are shown in Fig. 1. Physical quantities with solid arrows are primary data that should be transferred between codes as the law of causality, and those with shaded arrows govern feedback that can be neglected according to simulation circumstances.

We will present results of integrated fast ignition simulations that are performed by three codes, and demonstrate the capability of $FI^3$.

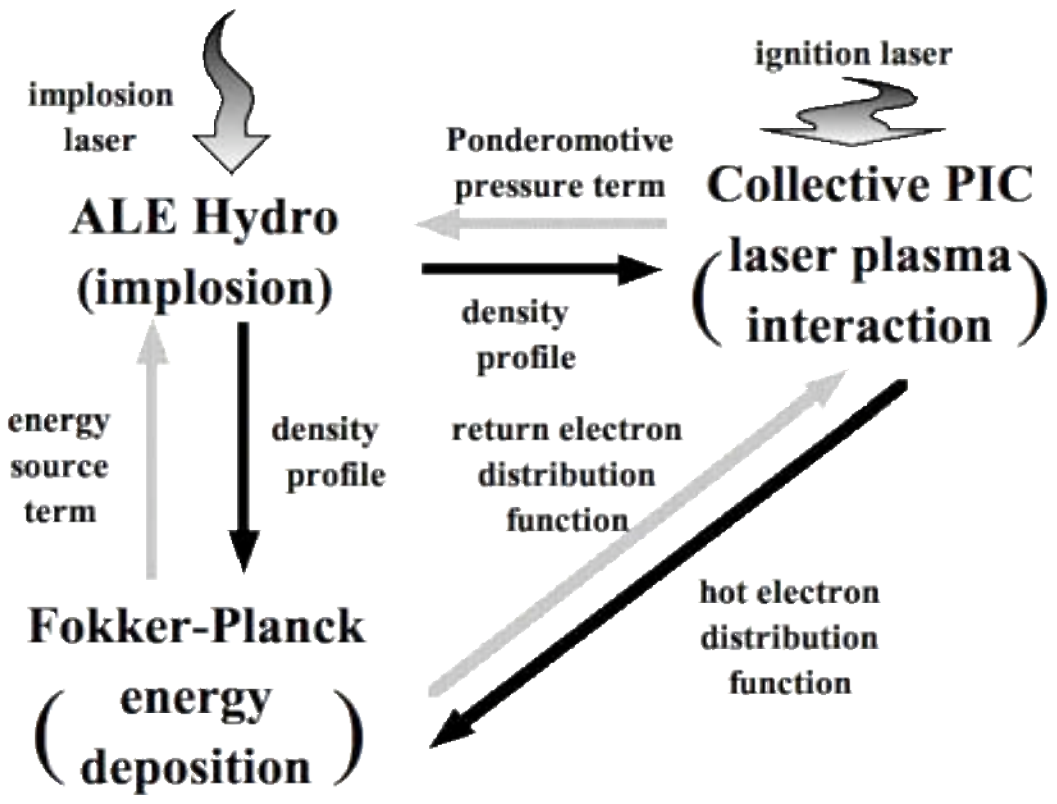

Fig.1 The concept of the code integration in Fast Ignition Integrated Interconnecting code project.

## 2. Integrated Simulations

### 2.1 Three Codes

In PINOCO, mass, momentum, electron energy, ion energy, equation of states, laser ray-trace, laser absorption, radiation transport, surface tracing and other related equations are solved simultaneously. Hydrodynamic is solved by ALE-CIP method, which can avoid the distorted mesh problem or rezoning/remapping problem of general Lagrangian based implosion codes. In the energy equations, diffusion equations that are the Spitzer-Harm type thermal transport model are solved using the implicit 9-point differencing with ILUBCG method. For the radiation transport, multi-group diffusion type model is installed in the code, in which we can use LTE or non-LTE (CRE) opacity database as table lookup.

Traditional PIC code requires an unrealistic huge number of particles to simulate such parameters and it is impossible to run the PIC code even on such massive parallel computers. To reduce number of particles and computations, we have introduced collective particles into FISCOF1. We statically get together many particles into one representative particle in overdense regions at an initial stage, and dynamically split a collective particle into many normal ones under appropriate conditions, but don't dynamically unify particles into one.

RFP has been developed for analysis of the fast electron transport and energy deposition processes in dense core plasma, which was coupled with a Eulerian hydro code to examine core-heating properties. In this code, a one-fluid and two-temperature hydro model treats cold bulk electrons and ions, and the RFP model treats the fast electrons generated by the ignition laser-plasma interactions. In the coupled RFP-hydro code, electromagnetic fields generated by fast electrons are evaluated by combining generalized Ohm's law, the Ampere-Maxwell equation, Faraday's law.

## 2.2 Distributed Computing Collaboration Protocol

We are trying to perform simulations that need to smoothly coordinate multiple codes, which communicate each other to integrate computations. For large-scale simulations, execution of each code is typically managed by NQS (Networking Queuing System; a job management system that assigns the queued task to idle processor by turns), and we cannot define when and where these codes are performed. Meanwhile, in order to communicate between two codes, both codes have to be simultaneously executed, and the sender code needs to know IP address of the receiver code. Thus communication between them is not necessarily possible due to NQS. In order to exchange data among codes, we therefore require a specific interface that has two features; sender and receiver codes do not have to be executed at the same time, and they can be performed at any machines. We have designed Distributed Computing Collaboration Protocol (DCCP) as a system that has the method of non-direct communication between codes. DCCP is a high-level protocol that uses TCP/IP, and builds up a unique communication system for exchanging data only. Especially collaboration of loosely coupled large-scale simulation codes is the main object of DCCP.

DCCP has two daemon processes, which are Arbitrator and Communicator, in the DCCP network to realize the feature as is described above. Communicator receives data that is sent from codes or transferred from other Communicators, deposits the data in its storage, and sends the data to codes that request the data or to other Communicators. In other words, Communicator actually performs data transmission instead of code itself. Arbitrator administrates all Communicators and codes in the DCCP network. Concretely speaking, it instructs Communicators to transfer data to other Communicators, and controls communications between codes. In the network with DCCP, one Arbitrator and multiple Communicators can exist. Fig.2 shows the configuration of DCCP. Control Signal is used for management of Communicators and codes, and data communication means transmission of data that are necessary for collaboration. Communicators use storage to temporarily store data for future usage.

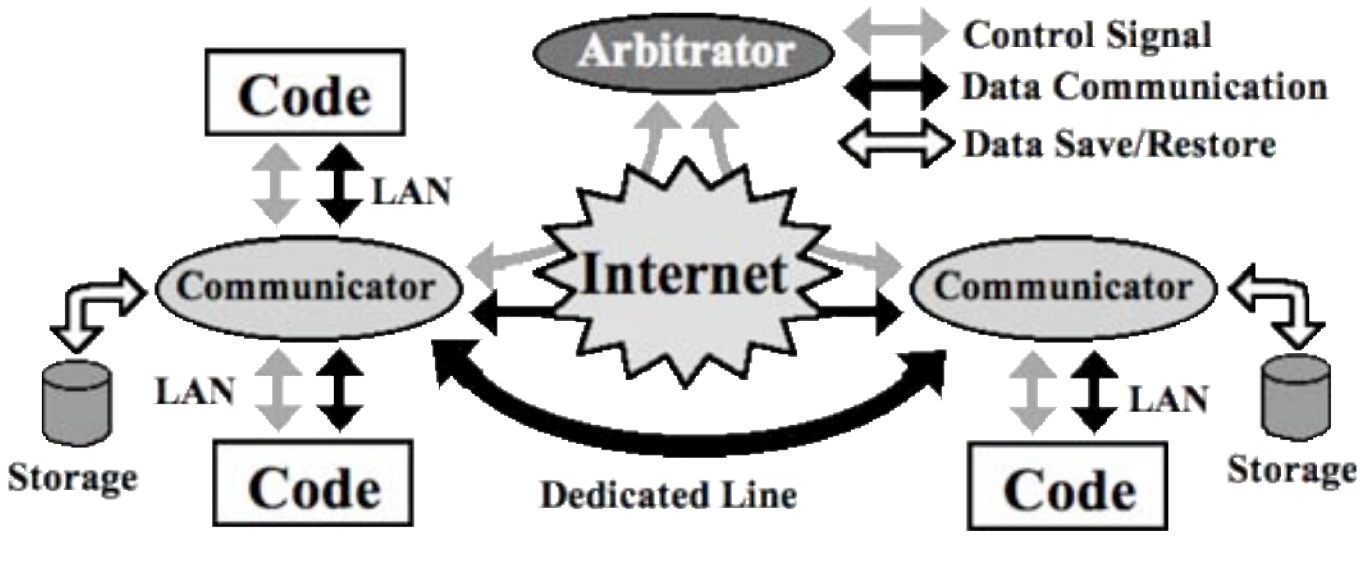

Fig.2 The configuration of the Distributed Computing Collaboration Protocol.

$\alpha$ version of DCCP was released in March 2004. In this version, the communication between computers with different Endian was invalid. As we supposed only a peer-to-peer communication and implemented a send/receive type data transfer, it was very inconvenient when one code wanted to send data to another two codes. The send/receive type was also inconvenient for parameter surveys with the same input data but different factors, because the data was deleted from the DCCP system once the receiver code got it and the sender code had to be executed just the same manner. Thus we fixed the Endian problem and introduced a multicast type communication, and released β version in Jun 2004. The DCCP version 1.0 was released in August 2004 with improvement of robustness, and Fortran calling interfaces of this version are summarized in Fig.3. The communication processes in the codes are very easily programmed using just five Fortran subroutines.

```
call DCCP_INITIALIZE (CodeName, RunName, Cnd)
call DCCP_SEND (DstCode, DstRun, N, Type, Data, Tag, Cnd)
call DCCP_UPLOAD (Ndst, DstCodes, DstRuns, N, Type, Data, Tag, Expire, Cnd)
call DCCP_RECEIVE (SrcCode, SrcRun, N, Type, Data, Tag, Timeout, Cnd)
call DCCP_FINALIZE (Cnd)
```

Fig. 3 Fortran calling interfaces of DCCP.

## 2.3 Estimation of Core Heating

In the first stage of $FI^3$ project, we demonstrated the integrated simulations for a cone-guided target and examined the implosion dynamics, the fast electron generation, and the core heating [9]. Time evolution of (a) the core heating rate and (b) the ion temperature, which are averaged over the dense core region ($\rho > 50 g/cm^3$), are shown for two different electron profiles, namely PIC source and mono-energy beam in Fig. 4. The PIC source is obtained by PIC simulations with the Gaussian pulse of $\lambda_L = 1.06$ [μm], $\tau_{FWHM} = 300$ [fs] and $I_L = 10^{20}$ [W/cm$^2$], and mono-energy beam is the Gaussian profile of $E_0 = 1$ [MeV], $\tau_{FWHM} = 300$ [fs]) and peak intensity of $3 \times 10^{19}$ [W/cm$^2$]. The electron profile obtained from the PIC simulation for $I_L = 10^{20}$ [W/cm$^2$] has almost the same intensity as the mono-energy beam, however, the average energy of electrons is excessively high so that the range is too long because the energy deposition rate per unit path length of a fast electron decreases with the electron energy. Thus the core heating rate obtained in the PIC source case is too small to raise the bulk plasma temperature, and we could not expect to heat the core in these simulations.

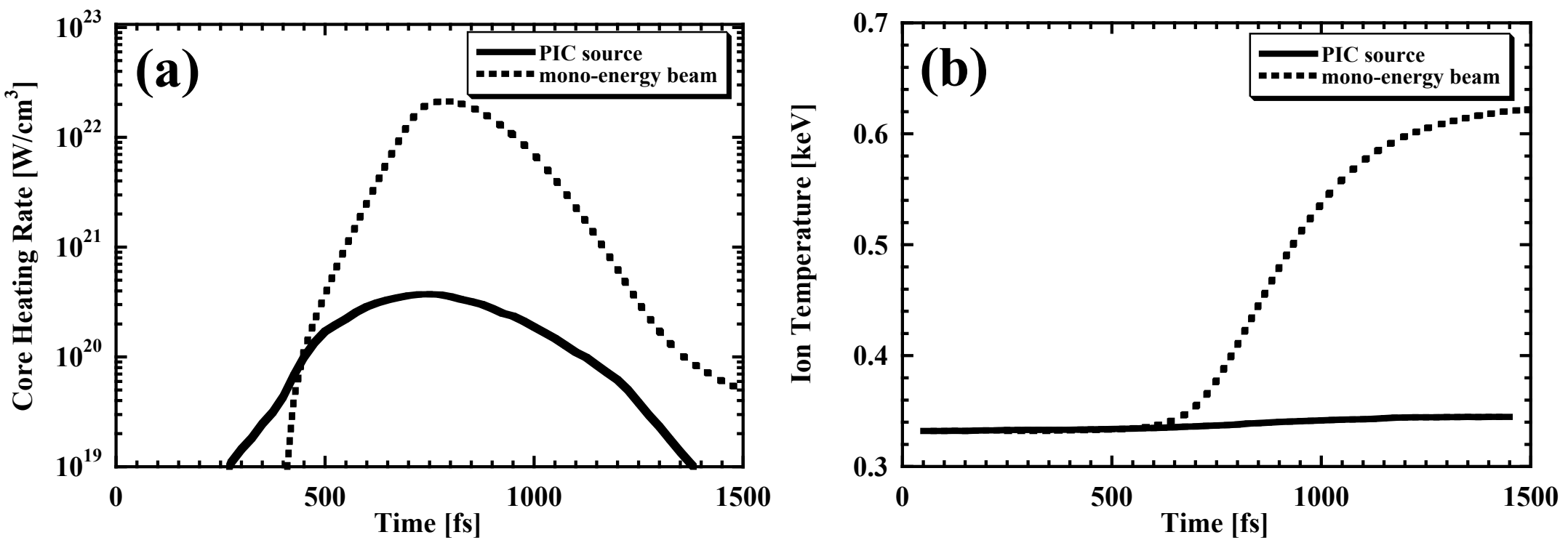

Fig. 4 Temporal profiles of (a) core heating rate and (b) ion temperature, which are averaged over the core region ($\rho > 50 g/cm^3$) for the PIC source (solid line) and the mono-energy beam (dash line).

## 3. Two-Stream Instability and Sloshing Electrons

Main reason why we could not get enough core heating with the PIC source was that fast electrons generated by the laser-plasma interaction are too hot to deposit their energy within the core. Thus we must consider another physics which are not taken into account in the previous integrated simulations. The ignition laser can directly interact with the cone wall like an oblique incident, and the vacuum heating could be important [10]. In addition, fast electrons that are generated at the cone wall travels along the wall and are accumulated at the cone tip [11]. We cannot treat those effects by FISCOF1 because of one-dimensionality, and are going to replace it with FISCOF2, which is a two-dimensional collective PIC code under development. We should also take care of recoiling electrons that are pulled back by the electrostatic potential and pass through the core many times with many chances to deposit their energy [12], and RFP should treat this recoiling effect. As a point of electron transport, it was reported by three-dimensional PIC simulations that the plasma return current experienced a strong anomalous resistivity due to diffusive flow of cold electrons in the magnetic perturbations, and the resulting electrostatic field leaded to an anomalously rapid stopping of fast MeV electrons [13]. Anomalous resistivity was also observed by three-dimensional hybrid simulations, and fast electron beam rapidly lost its energy accompanying with the filamentation [14].

In the cone-guided target, there is a density gap between the cone tip and the imploded plasma, and fast electrons must across this boundary to reach the core. So we model this density gap in FISCOF1 as follows: the preformed plasma is assumed to have exponential profile of the scale length 5 [μm] with density from $0.1n_c$ up to $100n_c$, where $n_c$ is the critical density. Behind the preformed plasma, the cone tip plasma is assumed as 10 [μm] width and $100n_c$, following the 50 [μm] compressed plasma with $2n_c$ (or $10n_c$). Fast electrons are expected to lose their energy when they across the density gap. The energy of fast electrons is evaluated at 10 [μm] rear of the density gap. The temporal profiles of the electron bean intensity and the time-averaged electron energy distribution for compressed plasma density of $100n_c$ (without the density gap), $10n_c$ and $2n_c$ with the Gaussian laser pulse of $\lambda_L$ = 1.06 [μm], $\tau_{FWHM}$ = 150 [fs] and $I_L = 10^{20}$ [W/cm$^2$] are shown in Fig. 5 (a) and (b), respectively.

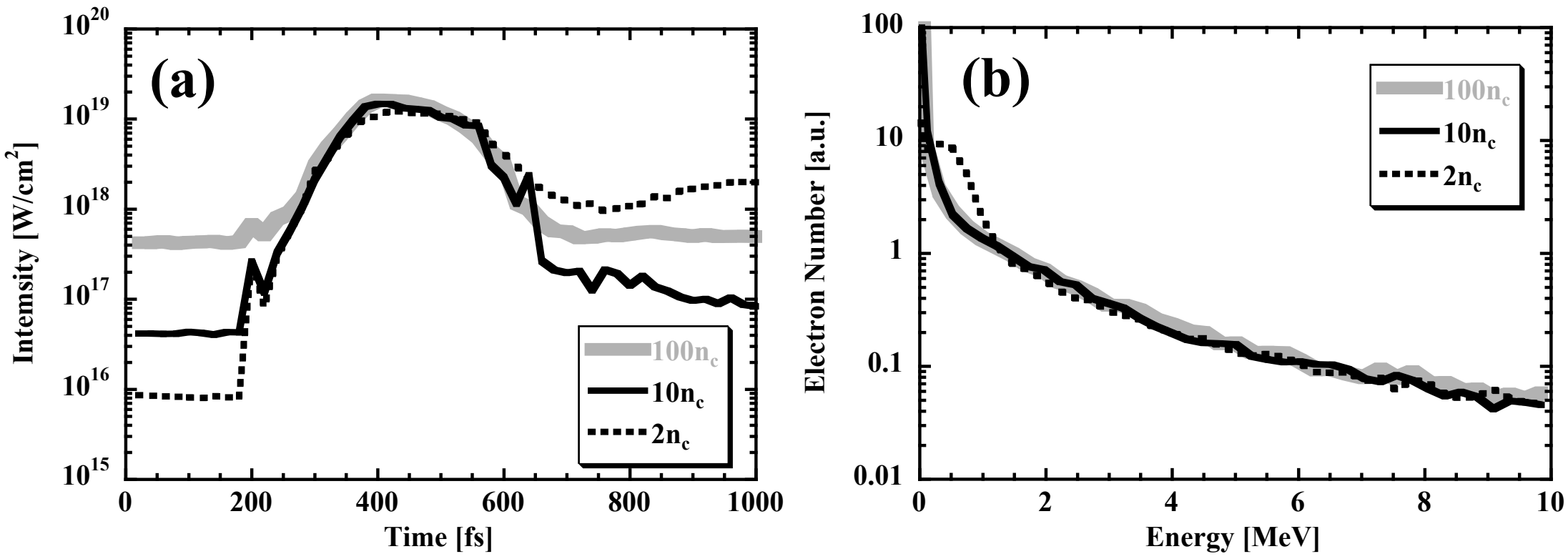

Fig. 5 Evaluation of fast electrons. The temporal profiles of (a) the electron bean intensity and (b) the time-averaged electron energy distribution for compressed plasma density of $100n_c$(gray solid), $10n_c$(black solid) and $2n_c$(black dash).

In case of $2n_c$, the electron beam intensity does not drop off even after the laser pulse vanishes, keeping higher than that of $100n_c$ case in Fig. 5(a), and sub-MeV electrons that are very efficient for core heating are clearly observed much more in Fig. 5(b). When the plasmas with different densities are adjacent, the contact potential is built up at the boundary. Amplitude of the contact potential could be assessed by the current conservation across the boundary as a function of the density ratio and was found to be the same order of the Boltzmann potential. Even the density ratio is 10,000, the amplitude is only $e\phi/T_e \sim 10$ and fast electrons can never be decelerated. Therefore there would be another mechanisms to generate sub-MeV electrons in case of $2n_c$. We estimated the fast electron current in the cone tip plasma in simulations and found that it reached ~ 0.1 x ($100n_c$ x $v_{te}$), where $v_{te}$ is the thermal velocity of bulk electrons and the electron temperature is set to 10 [keV] in simulations. To neutralize this current, the return current in the compressed plasma must be ~ 5 x ($2n_c$ x $v_{te}$). This means that all of bulk electrons in the compressed plasma must flow with 70% of the light speed because the ratio of the light speed to the electron thermal velocity is 7.15. This stream of bulk electrons excites a strong two-stream instability, in which bulk electrons are heated up. This is one reason for the obtained electron energy distribution in case of $2n_c$. If more of the fast currents are injected into the compressed plasma, they cannot be canceled due to limitation of the light speed, and the electrostatic field will be built up. Thus a portion of fast electrons which are generated by the laser-plasma interaction is confined inside the cone tip. On the other hand, the return currents are accelerated at the contact surface and injected into the cone tip, and finally reflected at the vacuum boundary. Therefore the cone tip is filled up with sloshing fast electrons, and some fractions of them are continuously released from there even after the laser pulse is dropped off, and expected to heat the core more than the case without the density gap.

It is noted that the density ratio of the cone tip to the compressed plasma is not important but the density difference between laser-produced fast electrons and bulk electrons of the compressed plasma is essential for this mechanism. When the density of bulk electrons is sufficiently higher than that of fast electrons, we can not expect this effect. It is also noted that the Weibel instability will occur instead of the two-stream instability in the multi-dimensional PIC simulations and the same effect for heating up the bulk will be expected.

## 4. Comparison with Experiments

We have performed $FI^3$ integrated simulations for the recent core heating experiment of cone-guided targets. Evaluated electrons by FISCOF1 in the previous section should be used as the time-dependent momentum distribution of fast electrons for RFP. But the duration of the ignition laser assumed in FISCOF1 simulations was shorter than that of experimental values (500~900fs). In addition, the laser-plasma interaction at the cone wall could not be included in the one-dimensional simulations, and the amount of fast electrons would be underestimation. Thus, the pulse duration and energy of the fast electron beam estimated above are so smaller than those expected in the experiments. To fit the source condition to the experiment, we made artificial corrections to the fast electron profile obtained by FISCOF1. First, the fast electron duration was stretched by a factor of five. Second, the fast electron intensity was adjusted so that the amount of "injected source electron energy" is equal to the laser energy $E_L$ (here, we assumed $E_L$ = 300J) in the experiments multiplied by the "energy coupling from laser to fast electron assessed by FISCOF1". We performed integrated simulations with corrected fast electron profiles, and we estimated the core heating properties. The temporal profiles of (a) core heating rates and (b) ion temperatures averaged over the dense core region (ρ > 50g/cm$^3$) for compressed plasma density of 2n$_c$(with the density gap) and 100n$_c$(without it) are shown in Fig.6.

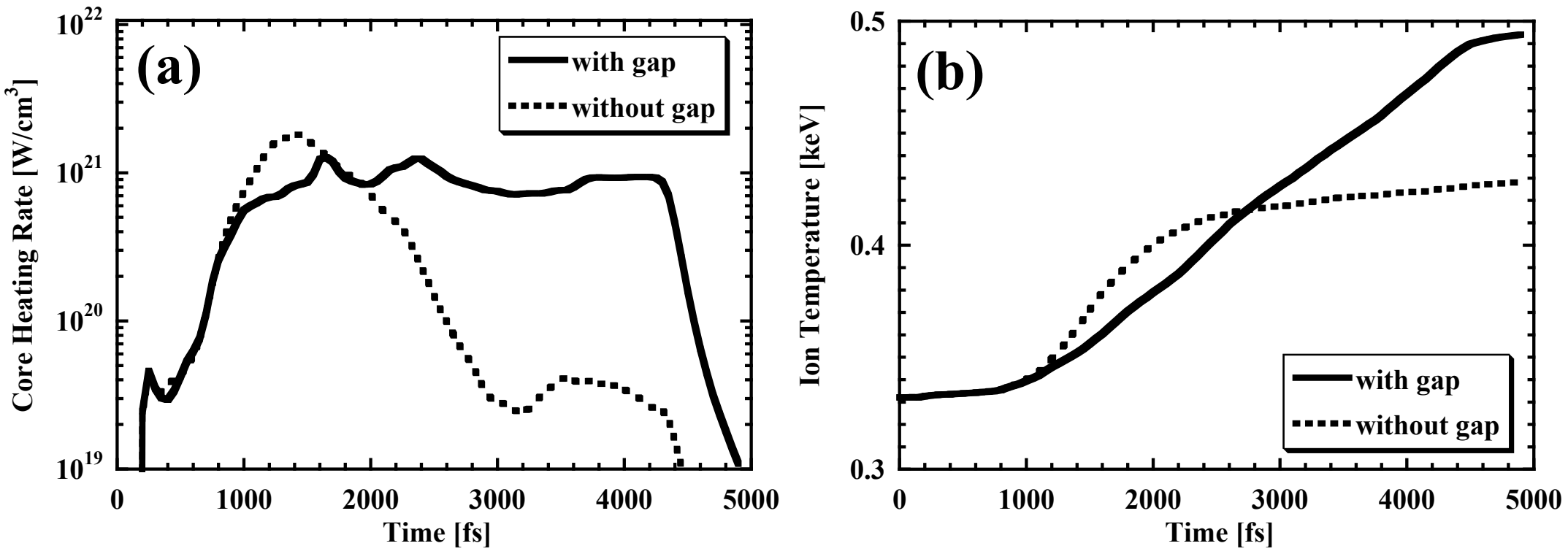

Fig. 6 The temporal profiles of (a) core heating rates and (b) ion temperatures averaged over the dense core region (ρ > 50g/cm$^3$) with (solid line) and without (dash line) the density gap.

The solid line shows the results for the case considering the density gap effect behind the cone tip, while the dash line represents the case neglecting this effect, respectively. In the case including the density gap effect, the electron beam intensity is lower than that of the neglecting case in the early stage (t < 2000 [fs]) because fast electrons are slightly decelerated by the electrostatic potential. As a result, the core heating rate with the gap is also lower and the ion temperature increases slowly. The core heating duration is, however, longer because the fast electrons with relatively low-energy, which are confined and sloshing inside the cone tip, are constantly delivered to the core after finishing laser irradiation. Thus the core is heated up from 0.33 [keV] to 0.49 [keV], ~ 69% higher increment by the density gap effect, but it is still lower than the temperature measured in the experiments, ~ 0.8 [keV].

## 5. Summary

We have shown the outline of the FI$^3$ project and the preliminary results of integrated simulations using three codes, the ALE hydro (PINOCO), the collective PIC code (FISCOF1), and the relativistic Fokker-Planck code (RFP). We have found that fast electrons generated by the laser-plasma interaction were too hot to efficiently heat the core. Considering the density gap at the contact surface between the cone tip and the imploded plasma, we estimated the fast electron profiles with one-dimensional PIC simulations. We found that the period of core heating became longer because bulk electrons were heated up to enough temperature to contribute for core heating by the two-stream instability, and some fractions of the fast electrons, which were confined and sloshing inside the cone, continuously exuded after vanishing the laser pulse and also heated the core.

Including the density gap effect, we performed FI$^3$ integrated simulations for the recent core heating experiment with cone-guided targets, and estimated the core temperature. The core was found to reach 0.49 [keV] with ~ 69% higher increment compared with the case neglecting the density gap effect. Even with the density gap effect, we cannot get core heating observed in the experiments by the present integrated simulations. As we introduced artificial corrections to realize conditions of the experiments, we should try to run the PIC code under the same parameters as the experiments. Each of the codes is also required to improve simulation model to fit more realistic situations.